\documentclass[12pt]{article}

\usepackage{newtxtext,newtxmath}

\usepackage{graphicx}

\usepackage[letterpaper,margin=1in]{geometry}

\renewenvironment{abstract}
	{\quotation}
	{\endquotation}

\date{}

\makeatletter
\renewcommand{\fnum@figure}{\textbf{Figure \thefigure}}
\renewcommand{\fnum@table}{\textbf{Table \thetable}}
\makeatother

\usepackage{scicite}

\usepackage{url}

\def\scititle{(In Review) A Comparative Hydrodynamic Characterization of the Flow Through Regular and Stochastically Generated Synthetic Coral Reefs Over Flat Topography \\ \large{[Under consideration at Coastal Dynamics 2025]} }
\title{\bfseries \boldmath \scititle}

\author{
	Akshay~Patil $^{1\ast}$,
	Clara Garc\'{i}a-S\'{a}nchez$^{1}$ \and
	\small $^{1}$ 3DGeoinformation Research Group, Delft University of Technology, The Netherlands.\and	
	\small $^\ast$ Corresponding author. Email: a.l.patil@tudelft.nl
}

\begin{document} 

\maketitle

\begin{abstract} \bfseries \boldmath
Coral reefs are vital to marine ecosystems, supporting biodiversity and driving nutrient cycling. Despite significant research on the interaction between surface waves and natural or artificial reefs, the turbulent flow dynamics within coral canopies remain poorly understood due to their intricate geometries. This study addresses this knowledge gap using a turbulence-resolving computation-al framework based on the volume-penalizing immersed boundary method (vIBM). Comparing the serial, staggered, and stochastic arrangements of various coral roughness types we observe that massive corals and cylinders lead to a similar hydrodynamics response and the effect of dispersive stresses can introduce a large difference when stochastic coral reefs are considered. These observations highlight the importance of better understanding the hydrodynamics of complex coral reef geometries, emphasizing the need for further studies on this aspect of coral reef hydrodynamics.
\end{abstract}

\section{Introduction}

The hydrodynamics of coral reefs play a central role in their ecological functions and resilience. By dissipating wave energy, promoting nutrient cycling, and supporting marine biodiversity, coral reefs are indispensable to both natural ecosystems and human societies \cite{Lowe2015}. However, their utility is under severe threat from climate change and anthropogenic stressors \cite{Caldeira2003}. Protecting coral reefs requires a comprehensive understanding of their hydrodynamic processes and concerted global efforts to mitigate threats. A wide range of studies have been conducted to better understand the hydro-dynamic response of coral reefs under varying forcing conditions \cite{Reidenbach2006,Stocking2016,Monismith2006,Davis2021,Lowe2005,Hamilton2024,Lowe2015}, to list a few. These studies have provided a deeper insight into the hydrodynamics over coral reefs evidenced by the recent developments \cite{Hamilton2024,Ascencio2022,Jacobsen2024,Jacobsen2022}.  In this study, we further the understanding of hydrodynamics over coral reefs by studying the effect of regularly arranged and stochastically arranged coral reefs using a scale-resolving computational framework. The focus here is to better understand and quantify the effect of relatively complex coral reef morphologies (referring to their complex spatial arrangement) compared to typically used serial and staggered arrangements comprising of a single coral species placed in a repeating fashion.

\section{Methodology}

\subsection{Governing Equations and Computational Methods}
To model the flow through the synthetic coral canopies, we solve the non-dimensional Navier-Stokes momentum equations subject to the incompressibility constraint given by

\begin{equation}
    \partial_t u_i + \Gamma \partial_j u_j u_i = \Gamma \left( - \partial_i p + \frac{1}{Re_k^b} \partial_j \partial_j u_i \right) + \cos(t) \delta_{i1} + F_{\text{IBM}},    
\end{equation}

\noindent and

\begin{equation}
    \partial_i u_i = 0,
\end{equation}

\noindent where $x_i$ are the coordinate directions corresponding to the streamwise, spanwise, and vertical directions respectively, $t$ is time, $u_i$ is the velocity vector, $p$ is the pressure, $\delta_{ij}$ is the Kronecker delta, $\Gamma \equiv U_b k_c/\omega$ is the relative-roughness, $Re_b^k \equiv U_b k_c/\nu$ is the roughness-height based wave Reynolds number, and $F_\text{IBM}$ is the immersed boundary force. The non-dimensional form of the equations is obtained by using the maximum wave orbital velocity ($U_b$), maximum coral canopy height ($k_c$), wave frequency ($\omega$), and inertial pressure scaling ($p \equiv p^*/\rho_o U_b^2$), where $p^*$ is the dimensional pressure and $\rho_o$ is the density of the fluid, and $\nu$ is the kinematic viscosity of the fluid. The governing equations are solved using a second-order accurate finite difference method \cite{Patil2022} where all terms in the governing equations are discretized explicitly, and the time integration is carried out through a three-step Runge-Kutta scheme using the fractional-step algorithm \cite{Ferziger2020}. The complex roughness is introduced into the computational geometry using a highly scalable signed-distance-field (SDF) generator \cite{Patil2024} using a volume-penalizing Immersed Boundary Method (vIBM) \cite{Scotti2006}. Further details on the computational method can be found in \cite{Patil2022}.

\subsection{Simulation parameters}

\begin{figure}
    \centering
    \includegraphics[width=0.5\linewidth]{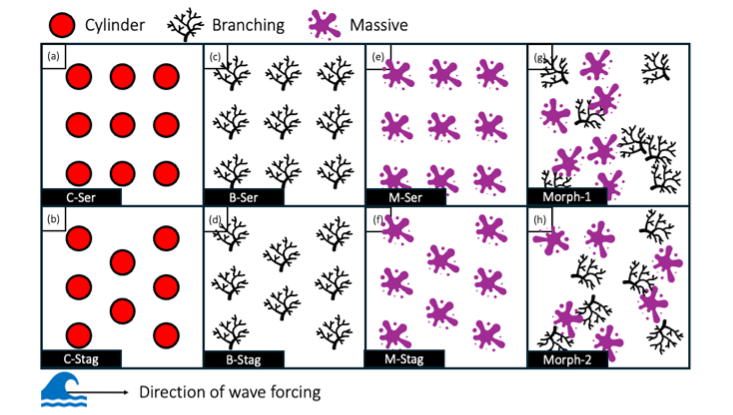}
    \caption{Morphological setup for the various simulations carried out in this paper along with the names of each of the scenarios. The case names are denoted at the bottom left of each panel, where, C stands for cylinder, B stands for Branching, M stands for Massive, Ser stands for serial, Stag stands for staggered, and Morph stands for stochastically generated morphology.}
    \label{fig:figure1}
\end{figure}

The coral geometries are obtained from the Smithsonian Institute coral repository \cite{Coral2023} and the stochastic coral reef was generated using two coral species, viz. \textit{Madrepora Formosa} and \textit{Pseudodiploria Strigosa} correspond to the branching and massive coral types that are typically used to represent coral geometries \cite{Hamilton2024}. In this paper, we compare the effect of morphological complexity of coral reef arrangements i.e., serial arrangement, staggered arrangement, and stochastic arrangement. Here morphology refers to the spatial arrangement of the individual corals that constitute the coral reef over a flat topography. Consequently, a total of 8 simulations are carried out, as shown in \ref{fig:figure1}, which details the arrangement of the cylinder and the coral geometries. Simulations Morph-1 and Morph-2 correspond to the stochastically generated coral reef using the two species of coral geometries. For cases with serial and staggered arrangements, the geometric centre-centre distance in both streamwise and spanwise directions is $S_x=S_y=k_c/1.75$ and the maximum coral canopy height $k_c=25.5 \delta_s$ where $\delta_s \equiv \sqrt{2\nu/\omega}$ is the Stokes’ wave boundary layer thickness and $\nu=10^{-6}$~$m^2/s$. The coral height-based wave Reynolds number is set to $Re_b^k=2533$ and the wave Reynolds number $Re_w \equiv U_b^2/(\omega \nu)=5000$. The domain size for the simulations is $20k_c \times 20k_c \times 2.5k_c$ in the streamwise, spanwise, and vertical directions, respectively. All the simulations were run on the Snellius supercomputer on the Genoa partition having 192 CPUs per node. Each simulation was run using 2 full nodes and required 65,520 CPU hours to simulate a total of 110 wave periods of which only the last 50 wave periods are used to calculate the statistics.

\section{Results and Discussions}

\begin{figure}
    \centering
    \includegraphics[width=0.8\linewidth]{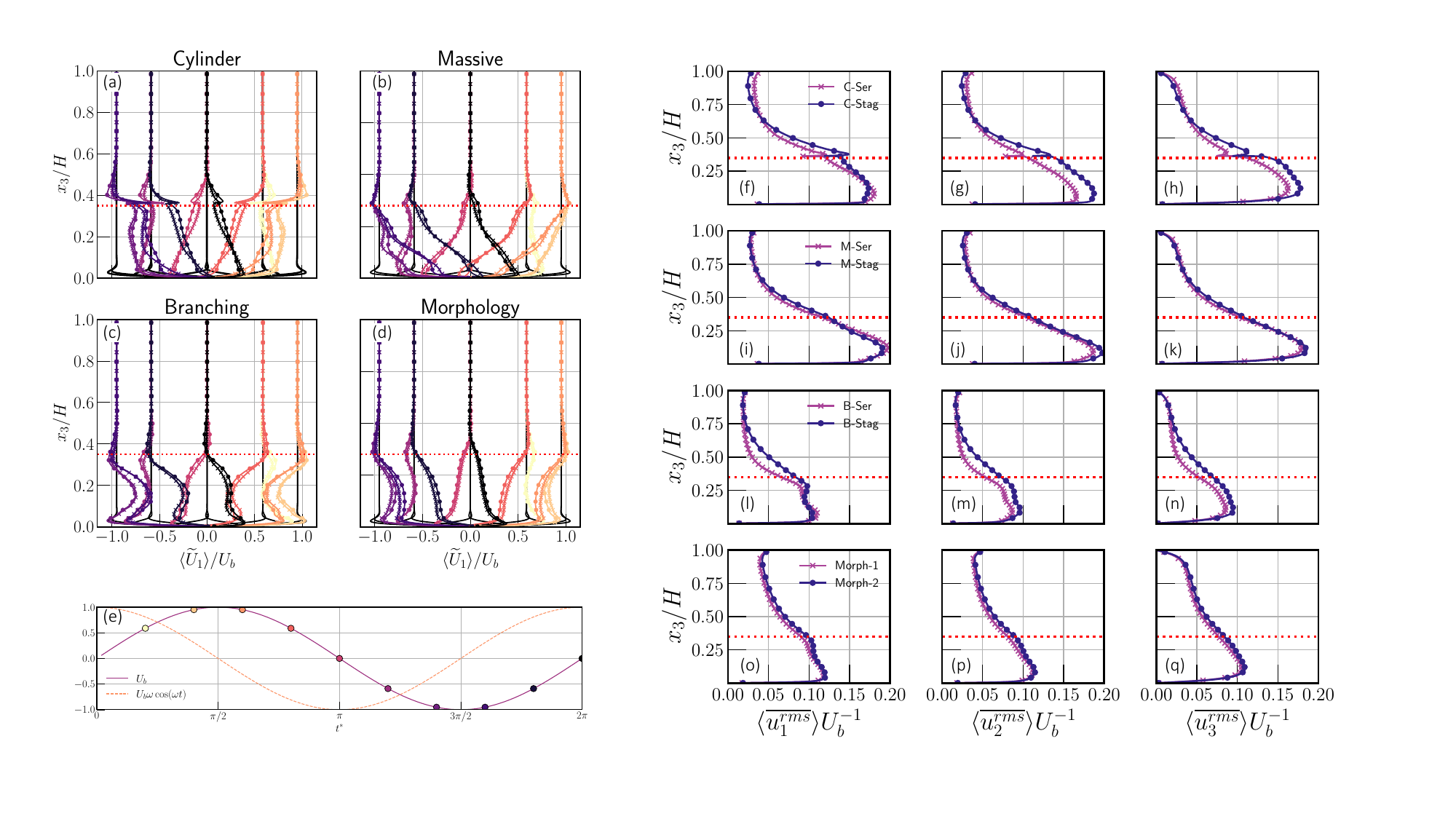}
    \caption{Plane- and Phase-averaged velocity streamwise velocity profile comparison for the four distinct geometries considered in this paper (panels a - d). Panels (f-q) compare the plane- and time-averaged root-mean-squared (rms) velocity components for all the cases simulated. Panel (e) shows the phase variations of the driving pressure gradient and the velocity with the filled circle markers denote the locations where the data in Panels (a-d) is presented. In Panels (a-d) the line markers correspond to those presented in panels (f-q). The horizontal, red-dashed line marks the location of the top of the coral reef roughness elements for all the cases.}
    \label{fig:figure2}
\end{figure}

Using a triple decomposition of the velocity given by

\begin{equation} \label{eq:tripleDecomp}
    U_i (x_i,t) \equiv \widetilde{U}_i (x_i,\omega t) + u_i^{\prime} (x_i,\omega t)=⟨\widetilde{U}_i ⟩(x_3,\omega t)+ \hat{U}_i (x_i,\omega t)+ u_i^{\prime} (x_i,\omega t),    
\end{equation}

\noindent where the terms on the right-hand-side in order are the plane- and phase-averaged velocity, dispersive velocity, and the turbulent velocity components, respectively. As shown in Figure \ref{fig:figure2}, for all the cases discussed in this paper, there are no substantial differences observed between the serial and staggered arrangements for a given coral geometry. However, when comparing the branching and massive coral types, there is a substantial difference in the near-wall velocity, where the branching type coral for both arrangements much closely follows the flat-wall Stokes’ solution \cite{Stoke1800}. In contrast, the massive corals that act as large roughness elements are observed to undergo an appreciable reduction in the wave velocity for all wave phases considered in this paper. This can be mainly attributed to the difference between the frontal area density ($A_f$) of the two coral types where the massive coral has a relatively larger frontal area density when compared to the branching type. Figure \ref{fig:figure2} further elucidates the dependence of the coral type when considering the stochastically generated coral reef which consists of both the massive and branching coral types and is observed to have a mean response that is halfway between either of the coral types for all wave phases.

Furthermore, when the plane- and time-averaged response of the turbulent rms velocity components is compared, there is a clear difference between the large $A_f$ coral reef (cylinder and massive) when compared with the relatively small $A_f$ coral reefs (branching and morphology). For all the cases, it is observed that the staggered arrangement led to a suppression of the streamwise velocity component at the expense of the other two when compared to the serial arrangement. As expected, for both the stochastic morphologies considered, there is identical flow response with minor differences that can be attributed to the stochastic nature of the coral reef. This further supports the observations made by \cite{Hamilton2024} where the primary differences are observed for the cylindrical roughness cases that exhibit a sharp shear layer at the top. For all the cases, far away from the bottom wall, the rms velocity profiles agree with each other irrespective of the geometric arrangement except for the cylindrical roughness that shows minor differences.

\begin{figure}
    \centering
    \includegraphics[width=0.8\linewidth]{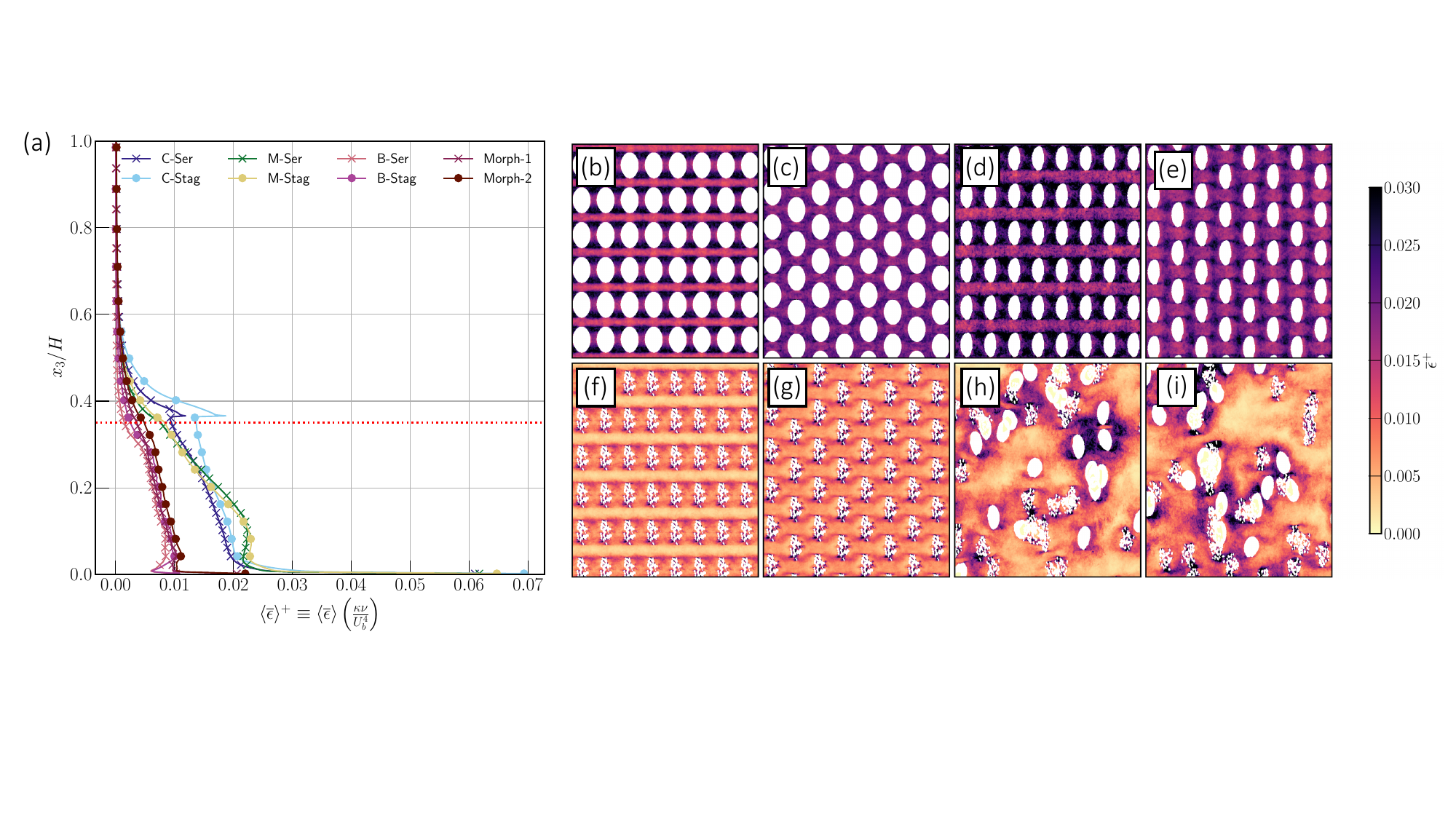}
    \caption{(a) Plane- and time-averaged turbulent kinetic energy dissipation rate profiles for all the cases. Panels (b-i) show time-averaged turbulent kinetic energy dissipation rate at $x_3/H = 0.2$ for all the cases discussed in this paper.}
    \label{fig:figure3}
\end{figure}

Figure \ref{fig:figure3}a presents the plane- and time-averaged turbulent kinetic energy (TKE) dissipation rate profiles for the various cases detailed in this paper. The similarities observed in the streamwise velocity are further exemplified when the TKE dissipation rate is compared. Specifically, cases C-Ser, C-Stag, M-Ser, and M-Stag are observed to exhibit a similar overall trend. Since the massive coral has a similar $A_f (x_3)$ as the cylinder, the hydrodynamic response is observed to be similar both at the large and the small scales. Interestingly, the effect of stochastic sampling of the coral geometry does not seem to have a large impact on the overall dynamics and can be modelled using branching type coral reefs, as they exhibit similar hydrodynamic response. As seen in Figure \ref{fig:figure3}h-i, the TKE dissipation rate is highly localized in the vicinity of large clusters of corals, while there are large patches of the flow domain where the TKE dissipation rate is small in magnitude. Overall, there is a strong influence of the $A_f$ on the small-scale dynamics of the flow suggesting that the dispersive stresses arising out of the triple-decomposition described in Equation \ref{eq:tripleDecomp}, can introduce strong heterogeneity in the flow leading to highly localized dissipation of the TKE. Consequently, further work is needed to accurately quantify and characterize the dispersive stress contribution in such environmental flows.

\section{Conclusions}
In this paper, we studied the similarities and differences in the hydrodynamic response of synthetically generated coral reefs using organized and stochastic arrangements of individual corals. Our work suggests that cylindrical roughness elements can act as adequate surrogates for massive coral type. Additionally, we observed large differences in the hydrodynamics response between the branching and massive corals primarily due to the differences in the frontal area density between the two. We also observed that using a stochastically generated coral reef with an equal mix of branching and massive coral types can provide similar results in the mean hydrodynamic response when compared to the staggered arrangement of identical branching-type corals. A principal finding in this work was the observation of localized zones of TKE dissipation in the stochastic coral reefs motivating further investigations into better characterizing the dispersive stress contribution to the overall dynamics of the flow. 
\paragraph*{Acknowledgments}
This work made use of the Dutch national e-infrastructure with the support of the SURF Cooperative supplemented by the Dutch Research Council (NWO). This publication is part of the project “Multi-fidelity computational modeling of environmental fluid systems” with grant number 2024/ENW/01763969.


\clearpage 

%
\bibliography{science_template} 
\bibliographystyle{sciencemag}

\end{document}